\documentclass[a4paper,11pt]{article}
\pdfoutput=1 

\usepackage{jinstpub} 

\title{\boldmath Tips for Deciphering and Quick Calculation
of Radiation Spectra}

\author[a,b]{M. V. Bondarenco}
\affiliation[a]{NSC Kharkov Institute of Physics and Technology, 
1 Academic St., 61108 Kharkov, Ukraine}
\affiliation[b]{V.N. Karazine Kharkov National University, 
4 Svobody Sq., 61077 Kharkov, Ukraine}

\emailAdd{bon@kipt.kharkov.ua}

\abstract{Radiation spectra from ultra-relativistic electrons in
thin [$T\ll l_f(\omega)$] and thick [$T\gg l_f(\omega)$] targets are
discussed. The method of simplified averaging is described by
examples of Landau-Pomeranchuk-Migdal effect and radiation at doughnut scattering. General
infrared and ultraviolet asymptotic properties of radiation spectra
are discussed.}


\proceeding{N$^{\text{th}}$ Workshop on X\\
  when\\
  where}

\begin{document}
\maketitle
\flushbottom

\section{Introduction}

The relation between the motion of a fast electron and the spectrum
of electromagnetic radiation emitted by it is often rather intricate.
In practice it is yet encumbered by an interplay of volume and edge
effects (see \cite{Bond-edgeeffects,Bond-Shul-double-scat} and refs.
therein) and by the necessity to average over random elements of the
electron trajectory. Fortunately, one can devise various approaches
for simplification. Some of them are based on asymptotic analysis,
and the others on simplified averaging procedures. The present
article discusses a few such approaches.




\section{Infrared asymptotics up to NLO [$l_f(\omega)\gg T$]}

An illustrative example is the infrared factorization theorem
\cite{Jauch-Rohrlich-IR}, stating that the limiting value of the
radiation spectrum at $\omega\to0$ (when the photon formation length
$l_f(\omega)=2\gamma^2/\omega$ greatly exceeds the target thickness
$T$) depends solely on the final electron deflection angle
$v_{fi}=|\vec{v}_f-\vec{v}_i|$ (with $\vec{v}_i$ and $\vec{v}_f$
being the initial and final electron velocities  obeying
$|\vec{v}_i|=|\vec{v}_f|$):\footnote{We adopt the system of units,
in which the speed of light equals unity, and denote by $\omega$ and
$\vec{n}$ the photon frequency and propagation direction, by $e$ and $m$ the
electron charge and mass, and by $\gamma=\sqrt{1-v^2}$ its Lorentz factor,
corresponding to the relativistic energy $E=m\gamma$. }
\begin{equation}\label{BH-semicl}
\frac{dI_{\text{BH}}}{d\omega}=\frac{e^2}{(2\pi)^2} \int d^2n
\left|\frac{\vec{n}\times\vec{v}_f}{1-\vec{n}\cdot \vec{v}_f}
-\frac{\vec{n}\times\vec{v}_i}{1-\vec{n}\cdot \vec{v}_i}
\right|^2
 =\frac{2e^2}{\pi}\!\left(\frac{2+\gamma^2v_{fi}^2}{\gamma v_{fi}\sqrt{1+\gamma^2 v_{fi}^2/4}}\text{arsinh}\frac{\gamma v_{fi}}{2}-1\right)\!.
\end{equation}
Formula (\ref{BH-semicl}) is independent of the detail of the
electron motion inside the target. To envisage the spectrum behavior
for all $\omega$, one often interpolates between (\ref{BH-semicl})
and the result found in the approximation of a ``thick'' target (see
Sec. \ref{sec:quick-averaging}). To make this procedure more
accurate, however, it is worth taking into account also the
next-to-leading order (NLO) correction to approximation
(\ref{BH-semicl}):
\begin{equation}\label{dIdomega-O(omega)}
\frac{dI}{d\omega}\underset{\omega\to0}\simeq
\frac{dI_{\text{BH}}}{d\omega}+C_1\omega+\mathcal{O}\left(\omega^2\right),
\end{equation}
with \cite{Bond-LowOmegaCorrection}
\begin{equation}\label{C1-semicl}
C_1=-\frac{e^2}2  \int_{-\infty}^{\infty} dt
[\vec{v}(t)-\vec{v}_i]\cdot [\vec{v}_f-\vec{v}(t)].
\end{equation}
Physically, the correction $C_1\omega$ is related to a difference
between the time delay $v\tau-\left|\vec{r}(\tau)-\vec{r}(0)\right|$
for the actual trajectory and that for its angle-shaped
approximation. In contrast to the Low theorem \cite{Low}, here $C_1$
depends on the electron dynamics inside the target.

\begin{figure}[h]
\includegraphics{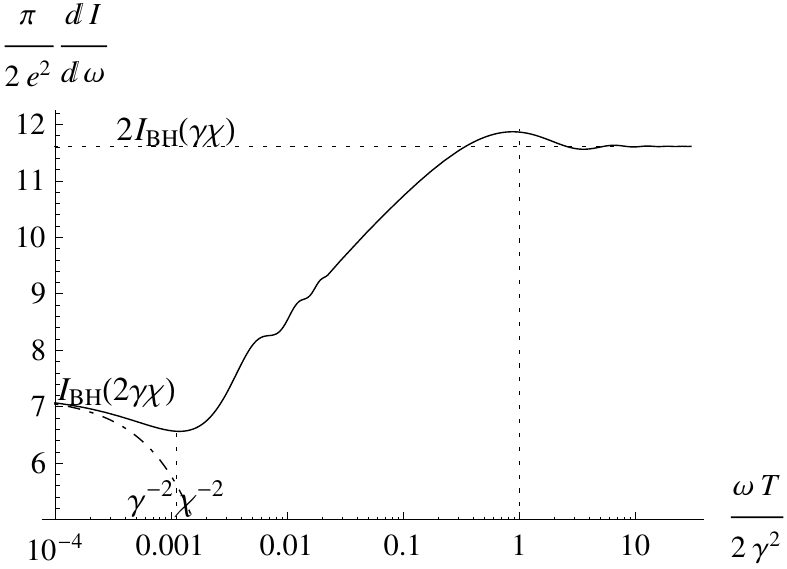}\hspace{2pc}%
\begin{minipage}[b]{14pc}\caption{\label{fig:Oscill} (Adapted from \cite{Bond-Shul-double-scat}). Spectrum of radiation at double scattering of an electron through
two equal successive elastic deflection angles
$\vec{\chi}_1=\vec{\chi}_2$. Due to the negative slope in the
origin, $C_1=-\frac{e^2}{2}\vec{\chi}_1\cdot\vec{\chi}_2
(t_2-t_1)<0$, the spectrum is non-monotonous at low $\omega$.}
\end{minipage}
\end{figure}

From Eq. (\ref{C1-semicl}) one infers that for monotonous electron deflection, $C_1< 0$ (see Fig.~\ref{fig:Oscill}), for an amorphous target, $C_1=0$, whereas for an oscillatory motion within the target, $C_1> 0$. For example, in case of undulator radiation, when the force acting on the electron has the form $\vec{F}_{\perp}(t)=\vec{F}_0\cos\frac{2\pi t}{T_1}$ within an interval $0<t<NT_1$, where $N\gg1$ is the number of oscillation periods of length $T_1$ each,
\begin{equation}\label{C1/NT1}
\frac{C_1}{NT_1}\underset{N\to\infty}\simeq
e^2\left(\frac{F_0 T_1}{4\pi E}\right)^2.
\end{equation}
At its application, it is worth noting that the correction
(\ref{C1-semicl}) is insensitive to non-dipole radiation effects.
Thus, relation (\ref{C1/NT1}), well known for dipole undulators,
must hold as well for wigglers, where the radiation spectrum is more
sophisticated.

\section{Radiation in thick targets [$l_f(\omega)\ll T$]. Quick averaging}\label{sec:quick-averaging}

For thick targets, when most of the photons are generated deeply
inside the target, it may be justified to neglect edge effects entirely and deal with
the radiation yield per unit time:
\begin{equation}\label{dI-domegadt}
\frac{dI}{d\omega dt}
=\omega\frac{e^2}{\pi}\int_0^{\infty}\frac{d\tau}{\tau}\Bigg\{\!\!
\left(\gamma^{-2}+\frac12\left[\vec{v}(\tau)-
\vec{v}(0)\right]^2\right) \sin \omega
\left[\tau-\left|\vec{r}(\tau)-\vec{r}(0)\right|\right]
-\gamma^{-2}\sin\omega (1-v)\tau\Bigg\},
\end{equation}
where the argument of the first sine can be evaluated via:
\begin{equation}\label{vtau-|r2-r1|}
v\tau-|\vec{r}(\tau)-\vec{r}(0)| \simeq
\frac{v}{2\tau}\int_{0}^{\tau}ds_2 \int_{0}^{s_2}ds_1
\left[\vec{v}(s_2)-\vec{v}(s_1)\right]^2.
\end{equation}

When the spectrum (\ref{dI-domegadt}) has to be averaged over random
variables, to avoid multiple integrals, the
pre-factor $\left[\vec{v}(\tau)-\vec{v}(0)\right]^2$ and the phase
(\ref{vtau-|r2-r1|}) may be replaced by their averages and
heuristically inserted into Eq. (\ref{dI-domegadt}). Such an
approach was used by Landau and Pomeranchuk in their pioneering
paper on LPM effect \cite{LPM}. It can not be regarded as rigorous, but is
attractive by its simplicity, so it is curious to examine how
accurately it can work in practice. In fact, in the dipole limit,
its prediction coincides with the exact result, so the only
remaining question is about its validity in the opposite, highly
non-dipole regime. Let us consider two examples.

\subsection{Radiation in an amorphous target}\label{subsec:amorph}

In case of electron passage through an amorphous target, evaluating
$ \left[\vec{v}(\tau)-\vec{v}(0)\right]^2=\left\langle
\frac{d\chi^2}{d\tau}\right\rangle \tau$, $
v\tau-|\vec{r}(\tau)-\vec{r}(0)|=\frac1{12}\left\langle
\frac{d\chi^2}{d\tau}\right\rangle\tau^2$, and inserting them to Eq.
(\ref{dI-domegadt}), we get $ \frac{dI}{d\omega dt}
=\frac{dI_{BH}}{d\omega dt} \tilde\Phi
\left(\frac{3\omega}{\gamma^4\left\langle
\frac{d\chi^2}{d\tau}\right\rangle}\right)
$.
Here $\frac{dI_{BH}}{d\omega
dt}=\frac{2e^2}{3\pi}\gamma^{2}\left\langle
\frac{d\chi^2}{d\tau}\right\rangle$ is the Bethe-Heitler spectrum, while
the form factor reads
\begin{eqnarray}\label{F-LPM-def}
\tilde\Phi(\Omega_a)=\frac98-\frac18\left|3-\sqrt{i\pi\Omega_a}e^{i\Omega_a/4}\text{erfc}\sqrt{\frac{i\Omega_a}{4}}\right|^2,
\end{eqnarray}
where $\text{erfc}(z)=\frac{2}{\sqrt{\pi}}\int_z^{\infty} dte^{-
t^2}$ is the complementary error function.

\begin{figure}[h]
\includegraphics{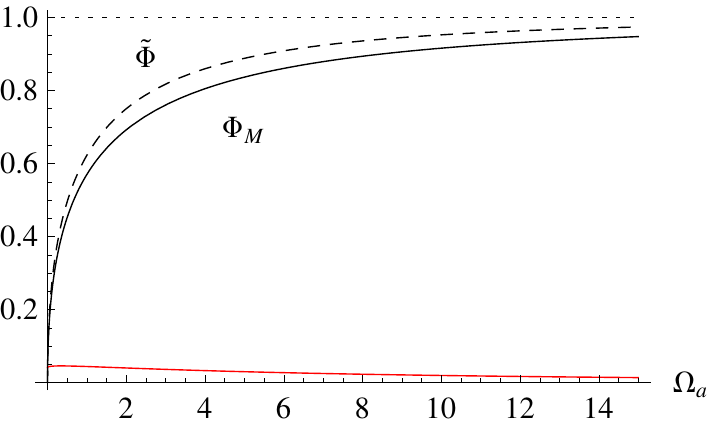}\hspace{2pc}%
\begin{minipage}[b]{16pc}\caption{\label{fig:Migdal-function} Comparison of Eq. (\ref{F-LPM-def}) (dashed curve) with Migdal's function $\Phi_M\left(\frac14\sqrt{\frac{\Omega_a}{3}}\right)$, $\Phi_M(s)=6s^2\left\{4\mathfrak{Im}\psi[(1+i)s]-\frac1s-\pi\right\}$ (solid curve). The red line shows the relative difference $\frac{\tilde\Phi-\Phi_M}{\tilde\Phi+\Phi_M}$. The vertical axis is in absolute units. }
\end{minipage}
\end{figure}

At $\Omega_a\to\infty$, form factor $\tilde\Phi$ tends to unity, saturating the Bethe-Heitler limit, which is natural since it corresponds to the dipole regime. On the other hand, in the infrared limit
$\Omega_a\to0$, $\tilde\Phi\simeq\frac34\sqrt{\frac{\pi\Omega_a}2}$.
Compared to the correct asymptotic behavior known from the Migdal's
theory, $\Phi_M\simeq\frac{\sqrt{3\Omega_a}}2$, it differs by a
factor of $\sqrt{\frac{3\pi}{8}}=1.085$, but for practical purposes,
such a difference may often be neglected (see
Fig.~\ref{fig:Migdal-function}).

\subsection{Radiation at doughnut scattering}

\begin{figure}[h]
\includegraphics{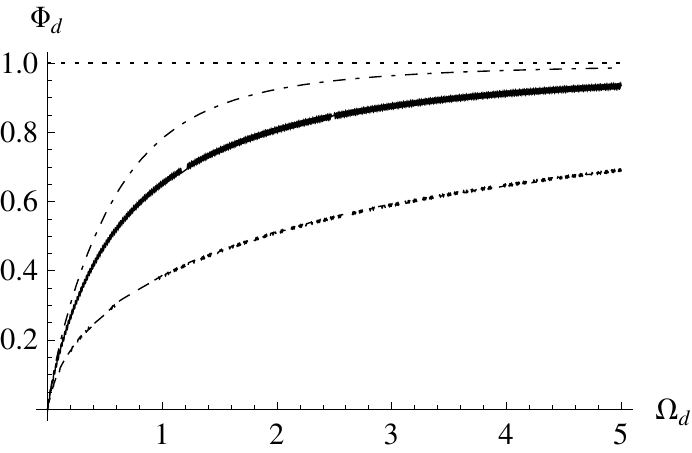}\hspace{2pc}%
\begin{minipage}[b]{16pc}\caption{\label{fig:Doughnut-rad} Behavior of form factor $\Phi_d=\frac{3\pi}{4e^2\gamma^2v_{\perp}^2 D}\frac{dI}{d\omega dt}$ evaluated by Eqs.~(\ref{doughnut-velocity-variance}), (\ref{doughnut-timedelay}), (\ref{dI-domegadt}). Dot-dashed curve, $\gamma v_{\perp}\to0$ [Eq.~(\ref{Phid0-def})]. Solid curve, $\gamma v_{\perp}=1$. Dashed, $\gamma v_{\perp}=3$.}
\end{minipage}
\end{figure}

A similar but more complicated example is radiation at electron
scattering on a family of aligned atomic strings in a crystal
(``doughnut scattering''). Assuming the strings to be mutually collinear and randomly distributed with uniform density in the transverse plane (which may be justified by the dynamical chaos in the electron
transverse motion), the kinetics of the electron multiple scattering
on the strings may be described by Fokker-Planck equation for the
probability distribution $f(\phi,t)$ in azimuthal angles $\phi$
between the velocity vectors relative to the string direction:
\begin{equation}\label{diff-eq-phi,r}
\frac{\partial f}{\partial \tau}=D\frac{\partial^2 f}{\partial \phi^2}.
\end{equation}
Here $D$ is the angular diffusion rate proportional to the string density and scattering strength. Solving Eq. (\ref{diff-eq-phi,r}) with the initial condition
$f(\phi,\vec{r}_{\perp},0)=\delta(\phi)\delta(\vec{r}_{\perp})$, we
get
\begin{equation}\label{doughnut-velocity-variance}
\left\langle [\mathbf{v}_{\perp}(\tau)-\mathbf{v}_{\perp}(0)]^2
\right\rangle=2v_{\perp}^2\left\langle1-\cos\phi
\right\rangle=2v_{\perp}^2(1-e^{-D\tau}),
\end{equation}
\begin{equation}\label{doughnut-timedelay}
v\tau -
\left|\vec{r}(\tau)-\vec{r}(0)\right|=\frac{v_{\perp}^2}{\tau}\int_0^{\tau}\!ds_2\int_0^{s_2}\!ds_1\left[1-e^{-D(s_2-s_1)}\right]=\frac{v_{\perp}^2}{D^2\tau}\!\left(1-D\tau+\frac{D^2\tau^2}2-e^{-D\tau}\right).
\end{equation}
The behavior of the spectrum obtained by plugging Eqs.
(\ref{doughnut-velocity-variance}), (\ref{doughnut-timedelay}) to Eq.
(\ref{dI-domegadt}) is shown in Fig.~\ref{fig:Doughnut-rad}. It
basically complies with experimental data of \cite{Bak-dougnhut}.

Let us now assess the accuracy of the adopted approach at $\gamma
v_{\perp}\gtrsim1$. Note that it interpolates smoothly between the
infrared and ultraviolet limits, so it is natural first to examine
the spectrum behavior in those two extremes.

In the ultraviolet (UV) limit,
\[
\frac{dI}{d\omega dt}
\underset{\Omega_d\gg1}\simeq \frac{4e^2\gamma^2v_{\perp}^2 D}{3\pi}
\]
(essentially, a dipole behavior). In the infrared (IR) limit,
\[
\frac{dI}{d\omega dt}
\underset{\Omega_d\ll1}\simeq\frac{e^2\omega}2 v_{\perp}^2,
\]
which is $\propto v_\perp^2$, as well. Thus, in the latter limit it
must be exact under averaging, too.

To figure out the intermediate-$\Omega_d$ behavior of the spectrum parametrically depending on $\gamma v_{\perp}$,
note first of all that at $\gamma v_{\perp}\to0$,
\[
\frac{dI}{d\omega dt}
\simeq\frac{4e^2\gamma^2}{3\pi v_{\perp}^2 D}\Phi_{d0}\left(\frac{\omega}{2D\gamma^2}\right),
\]
\begin{equation}\label{Phid0-def}
\Phi_{d0}(\Omega_d )=\frac{3\Omega_d}{2}\left\{(1-2\Omega_d^2)\text{arccot}\Omega_d+\Omega_d\left[2-\ln(1+\Omega_d^{-2})\right]\right\}.
\end{equation}
On the other hand, at $\gamma v_{\perp}\to\infty$, it tends to be proportional to $\tilde\Phi(3\Omega_d/\gamma^2 v_{\perp}^2)$, with $\tilde\Phi$ defined by eq. \ref{F-LPM-def}. At $\gamma v_{\perp}\sim1$, it spans intermediate values (see Fig.~\ref{fig:Doughnut-rad}). Since it works well in both extremes and interpolates smoothly between them, we may hope it to be numerically acceptable everywhere, thus giving a simple theory of radiation at doughnut scattering.



\section{Scaling in uniform media. LO and NLO IR and UV asymptotics}

Asymptotic behavior of radiation spectra at $\omega\to0$ and
$\omega\to\infty$ is chained to behavior of the correlators
$\left[\vec{v}(\tau)- \vec{v}(0)\right]^2$ and $v\tau -
\left|\vec{r}(\tau)-\vec{r}(0)\right|$ correspondingly at
$\tau\to\infty$ and $\tau\to0$. To generalize, suppose that the particle
motion in a uniform medium obeys a scaling law with an arbitrary index:
\begin{equation}\label{n-def}
\left\langle\left[\vec{v}(\tau)- \vec{v}(0)\right]^2\right\rangle=c_v\tau^n.
\end{equation}
Substitution thereof to Eq. (\ref{vtau-|r2-r1|}) yields
\begin{equation}\label{2(n+1)(n+2)}
\left\langle v\tau - \left|\vec{r}(\tau)-\vec{r}(0)\right|\right\rangle
=\frac{c_v}{2\tau}\int_{0}^{\tau}ds_2 \int_{0}^{s_2}ds_1
\left(s_2-s_1\right)^n =c_r\tau^{n+1},
\end{equation}
with $c_r=\frac{c_v}{2(n+1)(n+2)}$. For synchrotron radiation,
$n=2$, while for LPM effect, $n=1$. For doughnut scattering,
$n\simeq1$ for $t\ll D$, whereas $n\to0$ for $t\gg D$.

Employing these correlators in integral (\ref{dI-domegadt}), one can derive asymptotic expansion of the spectrum in the limit $\omega\to0$:
\begin{equation}\label{IR-scaling-NLO}
\frac{dI}{d\omega dt} \underset{\omega\to0}\simeq
\frac{e^2\sin\frac{\pi n}{2(n+1)}}{2\pi
(n+1)}\Gamma\left(\frac{n}{n+1}\right)\frac{c_v\omega^{\frac1{n+1}}}{c_r^{\frac{n}{n+1}}}-\frac{e^2\omega}{2\gamma^2}\frac{n}{n+1}+\mathcal{O}(\omega^2)
\end{equation}
(with $\Gamma$ the gamma function). Its leading order (LO) term is independent of $\gamma$, thus being
radiophysical by nature (see \cite{Bond-Shul-double-scat}). On the
other hand, the NLO term is independent of the strength of the force
acting on the particle, and may be the same for targets made of different materials with similar atomic order (e.g., single crystals Si and Ge in the
same orientation, or amorphous Al and Au, etc.).

In the ultraviolet limit, the asymptotics of the spectrum reads
\begin{equation}\label{spectrum-scaling-highomega}
\frac{dI}{d\omega dt} \underset{\omega\to\infty}\simeq \frac{e^2\gamma^2}{\pi}\Gamma(n)\sin\left(\frac{\pi n}{2}\right) 
(c_v-4nc_r)\left(\frac{2\gamma^2}{\omega}\right)^{n-1}
+\frac{e^2}{2\gamma}\sqrt{\frac{\omega n}{\pi}}\mathfrak{Re}\left\{\frac1{\tau_0}
e^{-\frac{\omega\tau_0}{2\gamma^2}\frac{n}{n+1}}\right\},
\end{equation}
with
$\tau_0=e^{i\pi\left(1/n-1/2\right)}\left[2(n+1)\gamma^2c_r\right]^{-1/n}$.
Generally, it involves a power law [the first term in Eq. 
(\ref{spectrum-scaling-highomega})], but at $n=2$ (a smooth electron
trajectory) the coefficient at it vanishes, so the decrease turns to
the synchrotron-like exponential described by the second term in
(\ref{spectrum-scaling-highomega}).

\begin{figure}[h]
\includegraphics[width=82mm]{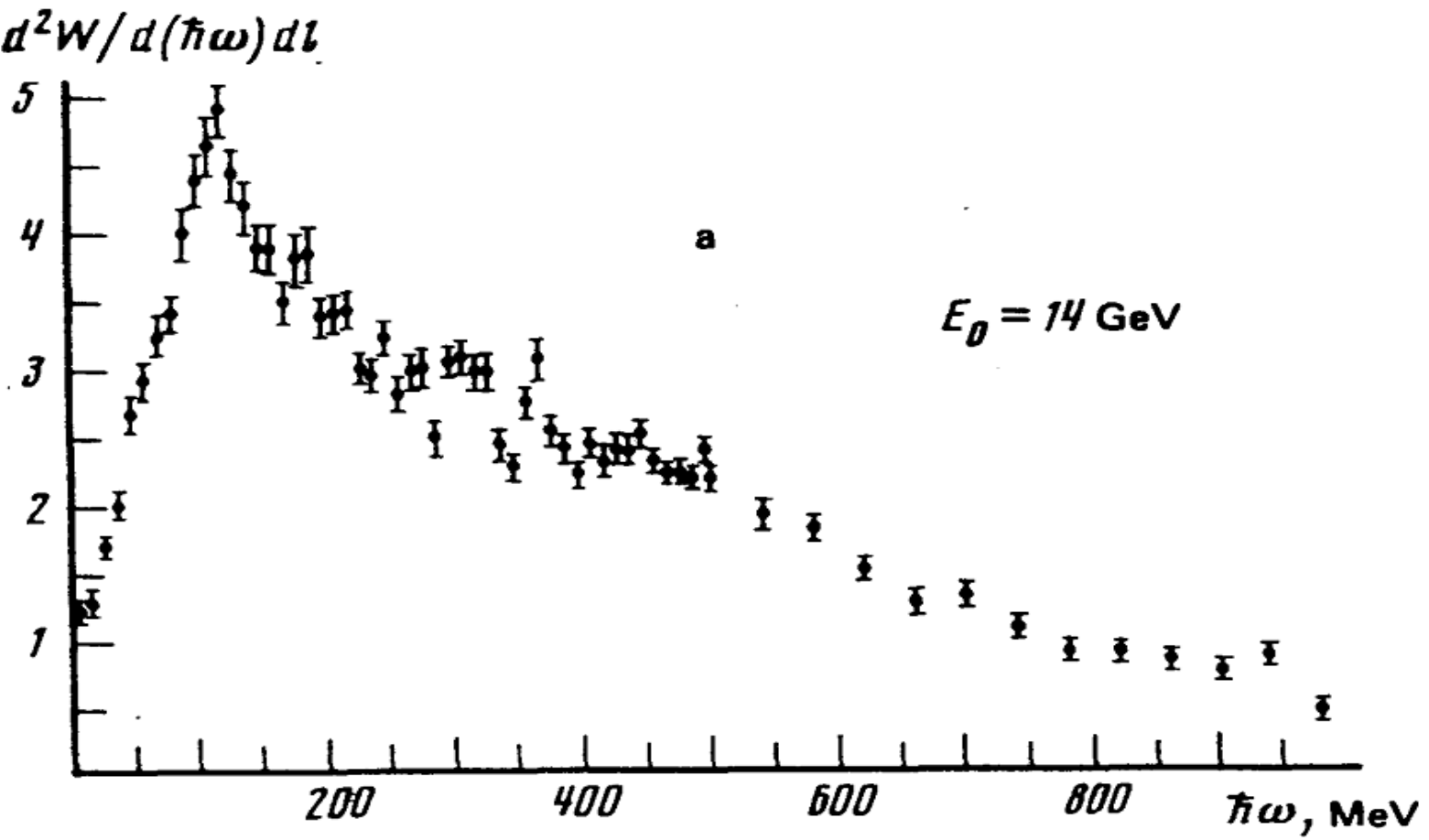}\hspace{2pc}%
\begin{minipage}[b]{14pc}\caption{\label{fig:Avakyan} (Adapted from \cite{Avakyan-channrad}) An example of experimental spectrum of channeling radiation (for crystal parameters see \cite{Avakyan-channrad}). Visually, the IR increasing and UV decreasing asymptotes may be extended to cover virtually the entire spectrum.}
\end{minipage}
\end{figure}

Using those rules with physically motivated values of $n$ at
$\tau\to\infty$ and $\tau\to0$, one can deduce asymptotics of the
spectrum correspondingly at $\omega\to0$ and $\omega\to\infty$. In
between, the spectrum is likely to interpolate smoothly. Sometimes,
the IR and UV asymptotes can cover almost the entire spectral region
-- see, e. g., Fig. \ref{fig:Avakyan}, where the two asymptotic
regimes seem to be accidentally valid up to their intersection point
(the channeling radiation peak).

An interesting question is whether there exist physical cases, in which
$n$ is non-integer. Computer simulations confirm this possibility
(cf., e.g., \cite{Greenenko}), but more studies are required.

\section{Summary}

In spite of diverse complications arising in practical radiation
problems, it is often possible to find ways for their material
simplifications. At $\omega\to0$, the radiation spectrum may be
highly non-dipole, but simple generic formulae in LO and NLO are
available. At large $\omega$, the radiation spectrum tends to be
more dipole, so if averaging is needed, it may be performed in a
simplified manner outlined in Sec.~\ref{sec:quick-averaging}. With
these tips, one can promptly connect radiation spectra with the
underlying electron dynamics; however, it is advisable to check such
predictions by more accurate numerical calculations.

\acknowledgments

This work was supported in part by the Ministry of Education and
Science of Ukraine (Project 0115U000473) and the
 National Academy of Sciences of Ukraine (Project CO-1-8/2017).




\begin{thebibliography}{19}

\bibitem{Bond-edgeeffects}
M.V.~Bondarenco, Mod. Phys. Lett. A \textbf{33}, 1850035
(2018).

\bibitem{Bond-Shul-double-scat}

M.V.~Bondarenco and N.F.~Shul'ga, Phys. Rev. D \textbf{95}, 056003
(2017); arXiv: 1703.05792.




\bibitem{Jauch-Rohrlich-IR}
F. Bloch and A. Nordsieck, Phys. Rev.  \textbf{52}, 54 (1937);

J.M. Jauch, F. Rohrlich, The Theory of Photons and Electrons, 1st
ed. (1955).


\bibitem{Bond-LowOmegaCorrection}
M.V. Bondarenco,  Phys. Rev. D \textbf{96}, 076009 (2017); arXiv: 1710.06459.


\bibitem{Low}
F.E. Low, Phys. Rev. \textbf{110}, 974 (1958).


\bibitem{LPM}

L.D.~Landau and I.Ya.~Pomeranchuk, Dokl. Akad. Nauk. SSSR
\textbf{92}, 535 (1953). 




\bibitem{Greenenko}

A.A. Greenenko, A.V. Chechkin, and N.F. Shul'ga,  Phys. Lett. A
\textbf{324}, 82 (2004).


\bibitem{Bak-dougnhut}
J. Bak \emph{et al.},  {\it Nucl. Phys.} B \textbf{302}, 525 (1988).

\bibitem{Avakyan-channrad}
R.O. Avakyan, I.I. Miroshnichenko, J.J. Murray, and T. Vigut, {\it
Sov. Phys. JETP} \textbf{55}, 1052 (1982).



\end{thebibliography}
\end{document}